\documentclass[fleqn,10pt]{wlscirep}

\usepackage{graphicx}
\usepackage{dcolumn}
\usepackage{bm}
\usepackage{color}
\usepackage{setspace}

\newcommand{\beq}{\begin{equation}}
\newcommand{\eeq}{\end{equation}}
\newcommand{\beqa}{\begin{eqnarray}}
\newcommand{\eeqa}{\end{eqnarray}}

\newcommand{\black}{\color{black}}

\title{Shortcut to adiabatic control of soliton matter waves by tunable interaction}

\author[1]{Jing Li}
\author[1]{Kun Sun}
\author[1,*]{Xi Chen}
\affil[1]{Department of Physics, Shanghai University, 200444, Shanghai, People's Republic of China}
\affil[*]{xchen@shu.edu.edu.cn}

\begin{abstract}
We propose a method for shortcut to adiabatic control of soliton matter waves in harmonic traps.
The tunable interaction controlled by Feshbach resonance is inversely designed to
achieve fast compression of soliton matter waves but within a short time, as compared to the conventional adiabatic compression.
These results pave the way to control the nonlinear dynamics for matter waves and optical solitons by using shortcuts to adiabaticity.
\end{abstract}
\begin{document}

\flushbottom
\maketitle
%
%


\section*{Introduction}
Tailoring matter wave in an accurate and fast way, including expansion/compression,
transport, and rotation \cite{Rabitz,Nature,EPL_david,Erik,AdreasNJP,Schaff3,Masuda}, in various time-dependent potential traps
becomes requisite with the applications ranging
from atom interferometry \cite{atominter}, metrology \cite{metrology} to quantum information processing \cite{walther}.
As usual, to guarantee the high fidelity, the ``slow" adiabatic process has been put forward and widely applied to prepare or
manipulate the quantum state of matter wave, avoiding
the final excitation. However, the slowness might destroy the control due to decoherence,
and energy consumption could be also problematic. Therefore, the techniques, sharing the
concept of ``shortcuts to adiabaticity" (STA) including invariant-based inverse engineering, counter-diabatic driving,
and fast-forward approaches, have been proposed to achieve adiabatic-like
control but within a very short time, see review \cite{Review}.

In particular, STA provides an efficient way to expand or compress matter waves without final heating in time-dependent harmonic traps \cite{PRL104,MugaJPB},
which has been experimentally demonstrated for cold atoms or Bose-Einstein condensates (BECs) atoms \cite{Schaff,Schaff2,Schaff3}.
The frictionless atom cooling in harmonic trap is also
interesting for thermodynamics \cite{tansient,yangyang}, and quantum heat engine or refrigerator \cite{AdC,Gong}.
Actually, STA solution can be inversely constructed when there exists a scaling solution for various systems. Therefore,
such shortcut control has been also extended from non-interacting single atom to interacting many-body systems \cite{David,Adolfo,Stringari,scale-invariance}.
Particularly, in order to implement shortcut to adiabatic control over BEC,
three main strategies are proposed by considering (i) the non-interacting limit;
(ii) a suitable driving of the interaction term; (iii) the Thomas-Fermi limit \cite{MugaJPB,Adolfo,Schaff2,Schaff3}. Alternatively,
the counter-diabatic driving, is also designed for speeded-up adiabatic manipulation of matter waves, in which
auxiliary nonlocal interactions and even potential painting technique are needed for different proposals \cite{AdolfoPRL}, for example, with applications of atomic loading in optical lattice \cite{MasudaPRL}.
Currently, counter-diabatic interaction has been further obtained to prevent the non-adiabatic transitions in nonlinear integrable systems based on the
equivalence between the equation of dynamical invariant and the Lax equation \cite{Takahashi}.

In this Report, we propose shortcut to adiabatic control of soliton
matter waves in harmonic traps, since the nonlinear dynamics of BEC is
of primary importance for coherent atomic optics, atom transport, and atom interferometry \cite{Science}. Instead of counter-diabatic driving \cite{Takahashi}, we apply the inverse engineering
to design the time-dependent nonlinearity, originating from controllable atom interactions, to compress the soliton rapidly.
Such shortcut reveals the existence of universal laws governing self-similar scale-invariance in this nonlinear system,
does not require changing trap frequency and adding a complementary potential.
To illustrate how the STA works for soliton compression, we first set the adiabatic reference by using
the hyperpolic tangent function of nonlinearity to modulate the atom interaction slowly. It seems that the long time required
to follow the adiabatic reference calculated by the analogy of perturbed Kepler problem.  In addition, the adiabatic result is not stable in asymptotical limit. To remedy such difficulties, we design inversely the smooth and feasible function of nonlinearity by choosing the same boundary conditions at initial and final times,
so that the soliton can be compressed from the same initial to final widths as adiabatic reference but within a short time.
Numerical simulation demonstrates that the compression of matter wave solitons can be accelerated at least by ten times, as compared to slow adiabatic compression \cite{Fkh_soliton}.
The results presented here are different from previous works \cite{PRL104,MugaJPB,Adolfo,AdolfoPRL}, in which the time-dependent trap frequency is engineered for achieving fast adiabatic-like (frictionless) evolution of cold atoms and the time dependence of nonlinearity is specified even for BEC atoms.
Here fast compression of the bright soliton for matter waves, as an extension, is proposed by tuning time-dependent nonlinearity, contributing from atom interaction, in terms of Feshbach resonance.

\section*{Results}



\textbf{Model and Hamilton}. It is well known that at zero temperature the dynamics of weakly interacting
Bose gas trapped in a time-dependent harmonic trap is well described by
Gross-Pitaevski equation (GPE). Here we consider the cigar-shaped harmonic
trapping potential with the elongated axis in the $x$-direction,
\begin{equation}  \label{eq1}
i\psi_t+\frac{1}{2}\psi_{xx}+g(t)|\psi|^2 \psi-V(x,t)\psi=0,
\end{equation}
where $\psi(x,t)$ is the wavefunction of the condensate, $g(t)$ is a time-varying function of nonlinearity.
$g>0$ ($g<0$) corresponds to the case of negative (positive) scattering length $a_s(t)$.
Considering physical potential, $V(x,t)=\omega^2[x-x_0(t)]^2$ is a real
function with trap frequency $\omega$. For convenience, all variables have been dimensionless in Eq. (\ref{eq1}). $%
t=\omega_{\bot}t^{\prime }$, $x=x^{\prime }/a_{\bot}$, $\psi$=$\sqrt{2|a_{s}|%
}\psi^{\prime }$, $V(x,t)=V^{\prime }/\hbar\omega_{\bot}$, $a_{\bot}=\sqrt{%
\hbar/m\omega_{\bot}}$ (transverse harmonic-oscillator length), $\omega_{\bot}$ is the transverse frequency, and \black $m$ is the
atomic mass. $g(t)=4\pi\hbar^{2}a_{s}(t)/m$ characterizes the strength of
interatomic interaction, which can be effectively modulated by s-wave
scattering length $a_{s}$ with Feshbach resonances in the experiment.

We begin with the scaling solution of Eq. (\ref{eq1}), which is the well-known bright solitary wave solution \cite{Lei}, represented as
\begin{eqnarray}
\label{anstz}
\psi(x,t)=A(t)\mathrm{sech}\left[\frac{x-\zeta(t)}{a(t)}\right]\exp\{i b(t)[x-\zeta(t)]^2+ic(t)[x-\zeta(t)]+i\phi(t)\}.
\end{eqnarray}
Here, $A\equiv A(t)$, $a \equiv a(t)$, $b\equiv b(t)$, $c\equiv c(t)$, $\zeta \equiv \zeta(t)$, and $\phi \equiv \phi(t)$ represent the amplitude, width, chirp,
velocity, center position and phase, respectively. They are all time
dependent and real functions. The normalization is $\int^{+\infty}_{-%
\infty}|\psi|^{2} dx=2 a A^{2}=2N$, resulting in $A=\sqrt{N/a}$.
Here two main parameters $a$ and $\zeta$ satisfy the following equations:
\beqa
\label{engeering}
\ddot{a}+4a\omega^2=\frac{4}{a^3 \pi^2}-\frac{4g(t)N}{\pi^2a^2},
\\
\ddot{\zeta}+4\omega^2[\zeta-x_0(t)]=0.  ~~~~
\label{transport}
\eeqa
Obviously, the width of wave packet $a$ is connected with time-dependent interaction $g(t)$
through Eq. (\ref{engeering}). Eq. (\ref{transport}) also provides the relation between the center of the mass of wave packet $\zeta(t)$ and trap center $x_0 (t)$.
Here we are concerned about the compression, instead of transport, so the harmonic trap is assumed to be static, $x_0(t)=0$, and the solution of Eq. (\ref{transport}) gives $\zeta(t) \equiv 0$.
In what follows we will play with Eq. (\ref{engeering}) for inverse engineering, thus achieving fast and perfect soliton compression by designing the atom
interaction appropriately, from initial state
\begin{eqnarray}
\label{initialstate}
\psi(x,0)=\sqrt{\frac{N}{a(0)}}\mathrm{sech}\left[\frac{x}{a(0)}\right]\exp\left[i \frac{\dot{a}(0)}{2a(0)} x^2\right],
\end{eqnarray}
to the final one
\begin{eqnarray}
\label{finalstate}
\psi(x,t_f)=\sqrt{\frac{N}{a(t_f)}}\mathrm{sech}\left[\frac{x}{a(t_f)}\right]\exp\left[i \frac{\dot{a}(t_f)}{2a(t_f)} x^2\right],
\end{eqnarray}
with the boundary conditions for $a$ and its derivative $\dot{a}$ at time edges, $t=0$ and $t_f$.

\begin{figure}[]
\begin{centering}
\scalebox{1.0}[1.0]{\includegraphics{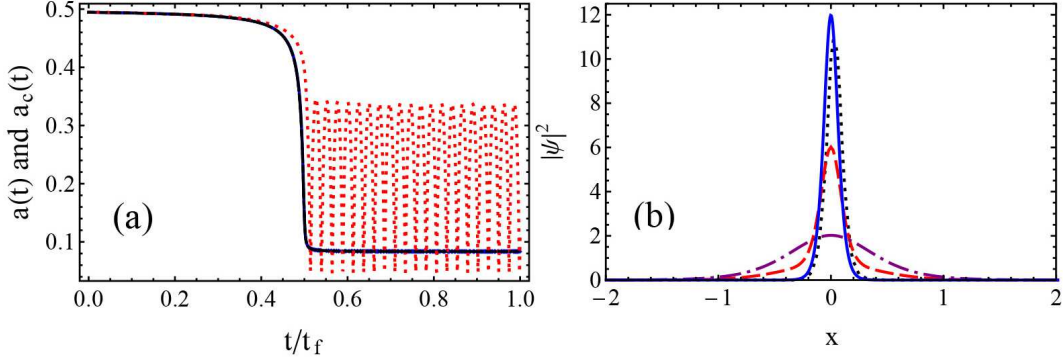}}
\caption{(Color online). (a) Evolution of soliton width $a(t)$ for adiabatic compression (solid blue)
and non-adiabatic compression (dashed red). The adiabatic reference $a_c(t)$ is undistinguishable from adiabatic case (dotted black).
(b) Wave function of adiabatic soliton compression with initial width $a(0)=0.494$ (dot-dashed purple) and final width $a(t_f)=0.0834$ (solid blue).
The numerical result (dotted black) is sightly different theoretical prediction (solid blue). The non-adiabatic compression (dashed red) is also compared.
Parameters: $A_s=10$, $s=1$, $\omega=0.04$, $N=1$, adiabatic case $t_f=100$, $s=1$, and non-adiabatic case $t_f=10$, $s=10$.}
\label{fig1}
\end{centering}
\end{figure}

\textbf{Adiabatic Reference}. First of all, we discuss the adiabatic reference of soliton compression before using shortcuts to adiabaticity in terms of
inverse engineering. To this end, we consider the analogy exists
between Eq. (\ref{engeering}) and the dynamical equation of a fictitious (classical) particle with position $x$,
as a perturbed Kepler problem \cite{Fkh_soliton,Fkh_optical}. The particle moves in the potential of the form,
\begin{equation}
U(t)= 2 a^2 \omega^2 - \frac{4g(t)N}{a\pi^2} +\frac{2}{\pi^2 a^2}.
\end{equation}
Newton's equation takes indeed the form of Eq. (\ref{engeering}) with certain kinetic energy $\dot{a}^2/2$. This analogy helps us to
achieve the adiabatic reference for further STA design. When $\omega =0$, $\partial U(t)/\partial a =0$ gives the minimum
point of potential
\begin{equation}
a_c (t) = \frac{1}{N g(t)}.
\end{equation}
On the other hand, when $\omega \neq 0$, the above equation gives the numerical solution $a_c (t)$ from the following expression:
\beq
\label{numerical}
a^4_c (t) \omega^2 \pi^2+ g(t)N a_c (t) - 1 =0.
\eeq
Assuming $\omega\ll 1$, we can obtain the analytical solution, by using perturbation, as \cite{Fkh_soliton}
\begin{equation}
a_c(t)=\frac{1}{Ng(t)} \left[1-\frac{\pi^{2}\omega^{2}}{g^4(t) N^4}\right],
\label{stable}
\end{equation}
which provides the minimum of perturbed Kepler potential of an effective particle. Thus, $a_c(t)$ in Eq. (\ref{stable}) provides the
adiabatic reference, when the interaction $g(t)$ is given, since trajectory $a_c (t)$ always gives the solution of non-adiabatic energy minimization \cite{yangyang}.
Without loss of generality, we choose the switching function of $g(t)$ as follows,
\beq
\label{gg}
g(t)=2+A_s\left\{1/2+1/\pi\arctan[s\pi(t-t_f/2)]\right\},
\eeq
where $t_f$ is final time, $A_s$ and $s$ are the control parameters, which determine the range and changing speed.
Once $g(t)$ is chosen, one can calculate the varying width of soliton by using Eq. (\ref{engeering}).
At the time edges, $t=0$ and $t_f$,  Eq. (\ref{gg}) gives
\beqa
g(0) &=& 2+A_s\left[1/2- 1/\pi\arctan(s\pi t_f/2)\right],
\\
g(t_f)&=& 2+A_s\left[1/2+1/\pi\arctan(s\pi t_f/2)\right],
\eeqa
which asymptotically trend to $g(0)=2$ and $g(t_f)= 2 +A_s$ when $s t_f \gg 1$.  Through Eq. (\ref{stable}),
we can further obtain the exact initial and final width $a_c (0)=0.494$ and $a_c(t_f)=0.0834$ for the parameters used in Fig. \ref{fig1}.
In Fig. \ref{fig1} (a) we compare the exact result calculated from Eq. (\ref{engeering})
with the adiabatic reference (\ref{stable}). To fix the same compression range, we choose $s t_f =100$.
For a long time, $t_f =100$, the evolution of $a(t)$ follows the adiabatic reference $a_c(t)$ exactly, while
the process is not adiabatic at all when $t_f =10$, so that the final width $a(t_f)$ does not coincide with $a_c (t_f)$.
In adiabatic case, final width $a(t_f)$ approaches asymptotically the adiabatic reference $a_c (t_f)$, see also Fig. \ref{fig1} (a).

Furthermore, numerical calculations are performed in Fig. \ref{fig1} (b) by solving the GP equation (\ref{eq1}) with time-dependent
nonlinear term $g(t)$, using the Crank-Nicholson scheme \cite{CN}. By using adiabatic protocol, we can compress the soliton
from the initial width $a (0)=0.494$ to the final one $a(t_f)=0.0834$. The results presented here almost coincide with the evolution of $a (t)$ in Fig.
\ref{fig1} (a). The slight difference results from the switching function (\ref{gg}), see the discussion below. However,
for non-adiabatic process, the wave function of soliton at final time $t_f$ is completely different from what
we expected from the adiabatic reference $a_c(t)$. Actually, multi-soliton could happen when the process is short.


\begin{figure}[]
\begin{centering}
\scalebox{1.0}[1.0]{\includegraphics{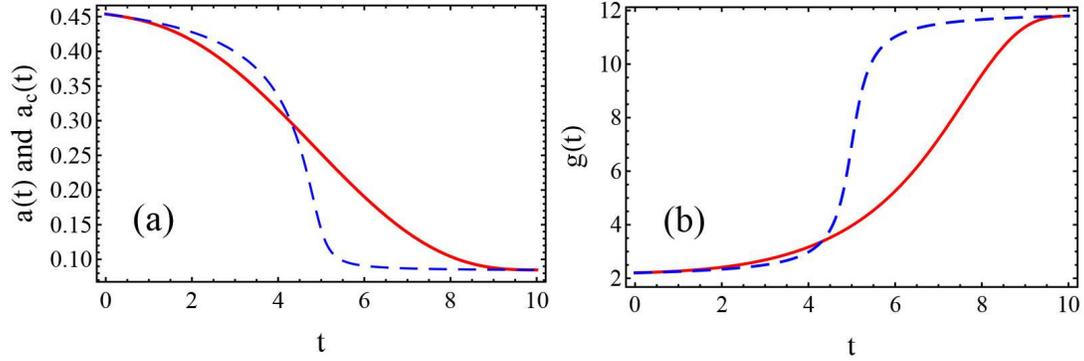}}
\caption{(Color online). Evolution of soliton  width $a (t)$  (a) and nonlinear parameter $g(t)$ (b) designed by STA (solid red) are compared with adiabatic references (dashed blue) by using hyperbolic tangent function Eq. (\ref{gg}) (solid blue). Parameters: STA $t_f =10$, $s=10$ and others are the same as those in Fig. \ref{fig1}.}
\label{fig2}
\end{centering}
\end{figure}

\begin{figure}[]
\begin{centering}
\scalebox{0.8}[0.82]{\includegraphics{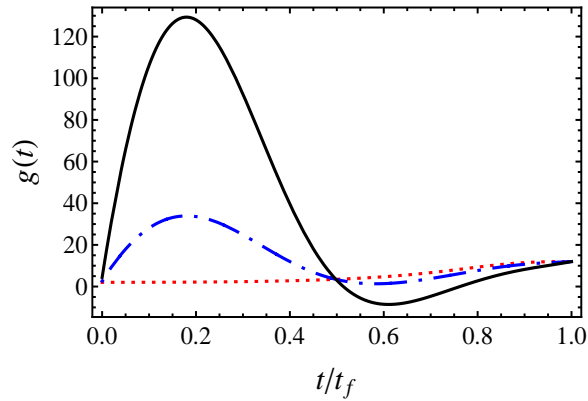}}
\caption{(Color online). Nonlinearity $g(t)$ versus $t/t_f$, where $s t_f=100$, solid black: $t_f=0.1$; dot-dashed blue: $t_f=0.2$, and dotted red: $t_f=10$. Other parameters are the same as
those in Fig. \ref{fig2}. }
\label{fig3}
\end{centering}
\end{figure}

\begin{figure}[]
\begin{centering}
\scalebox{1.0}[1.0]{\includegraphics{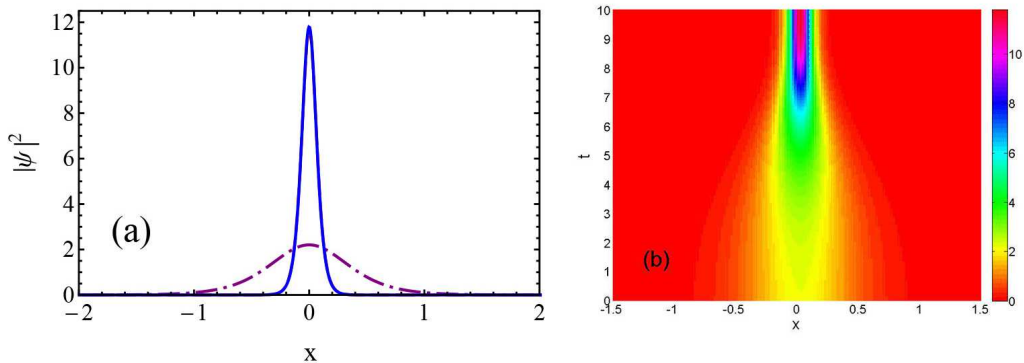}}
\caption{(Color online) (a)  Wave function of shortcut to adiabatic compression for soliton with initial width $a(0)=0.494$ (dot-dashed purple) and final width $a(t_f)=0.0834$ (solid blue).
The numerical result (solid blue) coincides exactly with theoretical prediction (dotted red), thus they are undistinguishable.
(b) Counterplot of evolution of soliton dynamics. Parameters are the same as those in Fig. \ref{fig2}.}
\label{fig4}
\end{centering}
\end{figure}

\begin{figure}[ht]
\begin{centering}
\scalebox{0.8}[0.8]{\includegraphics{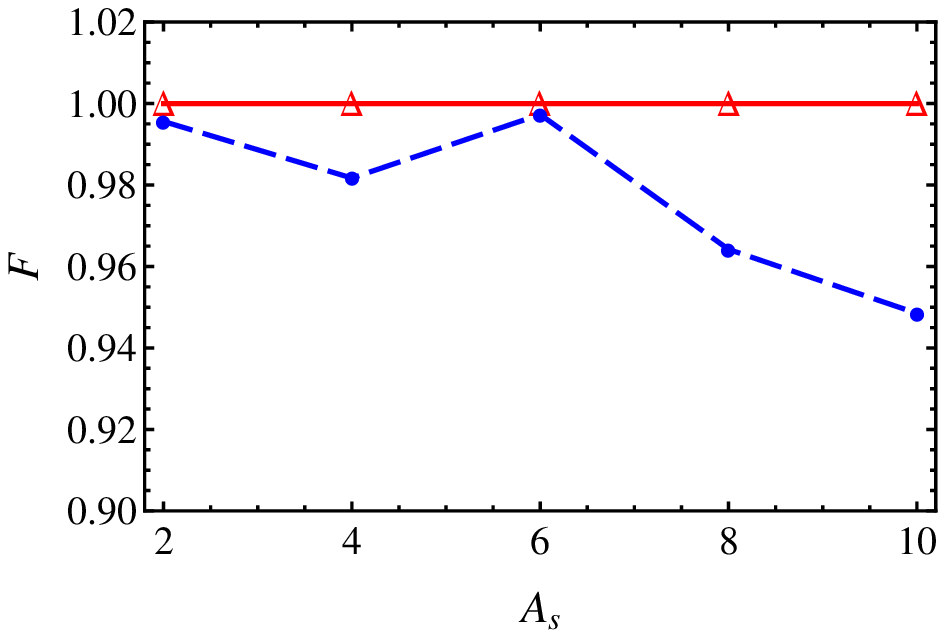}}
\caption{(Color online) Fidelity versus $A_s$ for adiabatic reference (dashed blue) and STA (solid red). Parameters are the same as those in Fig. \ref{fig1}.}
\label{fig5}
\end{centering}
\end{figure}

\textbf{Inverse Engineering}. Now we shall apply the auxiliary differential equation (\ref{engeering}) to construct STA protocol
in terms of concept on inverse engineering. To speed up the adiabatic process but achieving
the same target as the adiabatic reference, we have to set the following
boundary conditions at the time edges $t=0$ and $t=t_f$,
\beq
\label{BC-1}
a(0)=a_c(0),~~~ a(t_f)=a_c(t_f).
\eeq
In addition, more boundary conditions,
\beqa
\label{BC-2}
\dot{a}(0)=\dot{a}_c(0), ~~~ \dot{a}(t_f)=\dot{a}_c(t_f), \\
\label{BC-3}
\ddot{a}(0)=\ddot{a}_c(0), ~~~ \ddot{a}(t_f)=\ddot{a}_c(t_f).
\eeqa
are required to make  $\dot{a} (t)$ and $\ddot{a} (t)$ continuous at $t=0$ and $t=t_f$
through  Eq. (\ref{engeering}). From Eqs. (\ref{initialstate}) and (\ref{finalstate}),
the boundary conditions guarantee that during fast adiabatic-like compression of soliton the initial and final states should be the same as the adiabatic references with specified width $a_c(0)$ and $a_c (t_f)$.
Otherwise, the initial and final states are not stationary solution of GPE at initial and final times, so the STA design will be broken down. Once the boundary conditions are settled down,
we can use simple  polynomial ansatz $a(t)=\sum^{5}_{j=0}b_{j}t^{j}$ to interpolate
the function of $a (t)$, where the coefficients $b_{j}$ are solved by
using the boundary conditions Eqs. (\ref{BC-1})-(\ref{BC-3}). Finally the function
of nonlinearity, $g(t)$, can be constructed from Eq. (\ref{engeering}).
Since the boundary conditions are fixed in advance, the final result should be consistent
with the adiabatic reference for a arbitrarily short time. This guarantees how the STA works for fast adiabatc-like
compression of soliton. Noting that we consider the nonlinearity $g(t)$ as free parameter to control.
This is different from the previous shortcut design for BEC by using inverse engineering \cite{MugaJPB} and counter-diabatic driving \cite{AdolfoPRL}, where the non-interacting limit is assumed
or the specific driving of the interaction, $g(t)=1/a$, has to be imposed.

Figure \ref{fig2} (a) demonstrates that the evolution of $a(t)$ designed by STA does not
follow the adiabatic reference $a_c(t)$, but the initial and final values do coincide with the
adiabtic reference. This means the soliton can be compressed via STA.
In Fig. \ref{fig2} (b) the designed the function of nonlinearity, $g(t)$, are also depicted, in which the values of $g(0)$
and $g (t_f)$ are consistent with the values predicted by adiabatic reference, hyperbolic tangent function (\ref{gg}).
Mathematically, there is no limit for shortening the time by using STA, when the boundary conditions are satisfied.
However, physical constraint comes from the experimental implementation of time-dependent nonlinearity $g(t)$.
Since the time-dependent interaction is tuned by using Feshbach resonances in presence of magnetic or electric fields \cite{FRcontrol},
one can not modulate atom interaction as rapidly as we wish. In addition, when decreasing $t_f$ the maximum value of $g(t)$ becomes larger, see Fig. \ref{fig3}.
Supposing operation time is quite short, i.e.
$t_f \ll 0.2$, we will have negative value of $g(t)$ at certain time, and crossing resonance could induce atom loss.
Anyway, here we shorten the time by ten times as compared to adiabatic compression.
The smooth function of nonlinearity, $g(t)$, is reasonable for experimental implementation.

Figure \ref{fig4} (a) shows initial and final width of matter wave soliton, to demonstrate the shortcut to adiabatic compression.
Numerical calculation has a good agreement with the theoretical prediction.
STA can exactly compress the soliton to specific final width, as compared to adiabatic process.
In order to visualize it, we also show the dynamical evolution of soliton matter waves during the shortcut process, in Fig. \ref{fig4} (b).

\section*{Discussion}

Finally, we shall turn to the stability of STA with respect to different values of $A_s$. As mentioned above,
the amplitude of compression can be actually determined by the parameter $A_s$.
The fidelity, defined as $F= |\langle\psi^{ad} (x, t_f) |\psi (x, t_f) \rangle|^{2}$, is numerically calculated in Fig. \ref{fig5},
where $ |\psi (x, t_f)\rangle$ is the numerical result and $ |\psi^{ad} (x, t_f)\rangle$ is the ideal target state predicted from adiabatic reference.
Remarkably, by using STA protocol, we can compress the soliton to exact final width that we expected, when $t_f=10$ and $s=10$, and fidelity is always perfect, equal to $1$, showing
the advantage of STA on robustness. On the contrary, when $t_f=100$ and $s=1$, the process is adiabatic, and in general the fidelity is above $90 \%$, but there exists the oscillation
due to the switching function $g(t)$. This oscillation is relevant to hyperbolic tangent function for nonlinearity $g(t)$ (\ref{gg}),
similar to Allen-Eberly scheme in two-level system \cite{Vitanov}, and thus can be somehow suppressed by choosing other smooth switching function.
Additionally, we do not show the results for the parameters, $t_f=10$ and $s=1$ by using the switching function, $g(t)$ (\ref{gg}), since the fidelity
is far away from $1$, due to non-adiabatic transition. In fact, when decreasing $t_f$ further, the fidelity becomes imperfect, less than $1$.
This is due to the fact the values of $a_c(t)$ and its derivative at $t=t_f$ calculated from Eq. (\ref{numerical}) become inaccurate. One has
to modify the boundary conditions (\ref{BC-1})-(\ref{BC-3}) by replacing the numerical values calculated from  Eq. (\ref{numerical}),
instead of Eq. (\ref{stable}).



\section*{Conclusion}
In summary, we have proposed the shortcut to adiabatic control of soliton matter waves in harmonic trap by using inverse engineering.
The variational method used here provides new auxiliary differential differential equation (\ref{engeering}) for engineering inversely
the time-dependent nonlinearity $g(t)$, implemented by tunable Feshbach resonance. As an example, we first fix boundary conditions from adiabatic reference, design the function of nonlinearity $g(t)$
and finally achieve fast adiabatic-like compression. This is completely different from other STA protocols for BEC manipulation, in which
the time-dependent nonlinearity has to be imposed in both methods of inverse engineering and counter-diabatic driving.
The physical constraint on shortening time is discussed, and the time-optimal problem and other optimization
will be interesting for further investigation by combing STA and optimal control theory, see example in Ref. \cite{OCT}.
In addition, several extension could be done in the near future. For instance, one can transport the soliton matter wave \cite{Erik,AdreasNJP} simultaneously along with compression/expansion.
Optical soliton can be compressed in optical fibre by modulating gain and loss \cite{JOSA}.
These results presented here pave the way to study the fast and robust control of nonlinear dynamics in classical and quantum systems \cite{Takahashi,Fkh_optical}.

\section*{Methods}

The Lagrangian density
is \cite{Lagrang,Zoller}
\begin{eqnarray}
\label{Lagdensity}
L=\frac{i}{2}\left(\frac{\partial \psi}{\partial t}\psi^{*}-\frac{\partial
\psi^*}{\partial t}\psi\right)-\frac{1}{2}|\frac{\partial \psi}{\partial x}%
|^2  +\frac{1}{2}g(t)|\psi|^4-2\omega^2[x-x_0(t)]^2|\psi|^2.
\end{eqnarray}
Inserting ansatz Eq. (\ref{anstz}) into Eq. (\ref{Lagdensity}), we can calculate a grand Lagrangian by
integrating the Lagrangian density over the whole coordinate space, $%
\mathcal{L}=\int^{+\infty}_{-\infty} Ldx$. The Euler-Lagrange
formulas $\delta \mathcal{L}/\delta p=0$, where $p$ presents one of
the parameters $a$, $b$, $c$ and $\zeta$, give the following coupled
differential equations:
\begin{eqnarray}
\dot{a} &=& 2ab, \label{a} \\
\dot{b}&=&\frac{4}{a^4\pi^2}-2b^2-\frac{2g(t)N}{\pi^2 a^3}-2\omega^2, \\
\dot{\zeta}&=&c, \\
\dot{c}&=&-4\omega^2[\zeta-x_0(t)] \label{c}.
\end{eqnarray}
Note that $\delta \mathcal{L}/\delta \phi=0$ is tantamount to the
conservation of the norm of the wave function, which is the single dynamical
invariant of Eq. (\ref{anstz}), thus the absolute phase variable $\phi(t)$ plays no
role in the variational dynamics. Finally, the above coupled differential equations
(\ref{a})-(\ref{c}) become Eqs. (\ref{engeering}) and (\ref{transport}). Here we
choose hyperbolic secant ansatz, see Eq. (\ref{anstz}), the solution of nonlinear GPE, which is different from
the Gaussian one as the ground state in the linear limit (no interactions) \cite{Zoller}.


\section*{Acknowledgements}
We thank Yong-Ping Zhang's helpful discussions.
This work was partially supported by the NSFC (11474193),
the Shuguang program (14SG35), the Specialized Research Fund for the Doctoral Program (2013310811003),
the Program for Professor of Special Appointment (Eastern Scholar).

\section*{Author contributions statement}

J.L. carried out the theoretical and numerical calculations. K.S. and X.C. analyzed the results. X.C. conceived the idea and supervised the whole project. All authors wrote and reviewed the manuscript.

\section*{Additional information}
To include, in this order: \textbf{Accession codes} (where applicable); \textbf{Competing financial interests}: The authors declare no competing financial interest.

\end{document}